\documentclass[twocolumn,prb,showpacs,preprintnumbers,superscriptaddress,amsmath,amssymb]{revtex4}
\usepackage{graphicx}
\usepackage{dcolumn}
\usepackage{color}
\usepackage{bm}
\newcommand{\ud}{\,\mathrm{d}}

\DeclareMathOperator{\sech}{sech}

\begin{document}

\title{Electron scattering from domain walls in ferromagnetic Luttinger liquids}

\author{N.~Sedlmayr}
\email{sedlmayr@physik.uni-kl.de}
\affiliation{Department of Physics, University of Kaiserslautern,
D-67663 Kaiserslautern, Germany}
\author{S.~Eggert}
\affiliation{Department of Physics, University of Kaiserslautern,
D-67663 Kaiserslautern, Germany}
\affiliation{Research Center OPTIMAS, University of Kaiserslautern,
D-67663 Kaiserslautern, Germany}
\author{J.~Sirker}
\affiliation{Department of Physics, University of Kaiserslautern,
D-67663 Kaiserslautern, Germany}
\affiliation{Research Center OPTIMAS, University of Kaiserslautern,
D-67663 Kaiserslautern, Germany}

\date{today}

\begin{abstract}
  We study the properties of interacting electrons in a
  one-dimensional conduction band coupled to bulk non-collinear
  ferromagnetic order. The specific form of non-collinearity we
  consider is that of an extended domain wall.  
  The presence of ferromagnetic order breaks spin-charge separation
  and the domain wall introduces a spin-dependent scatterer active
  over the length of the wall $\lambda$. Both forward and backward
  scattering off the domain wall can be relevant perturbations of the
  Luttinger liquid and we discuss the possible low temperature phases.
  Our main finding is that backward scattering, while determining the
  ultimate low temperature physics, only becomes important at
  temperatures $T/J < \exp(-\lambda/\lambda_+)$ with $J$ being the
  magnetic exchange and $\lambda_+$ the backward scattering length
  scale. In physical realizations, $\lambda \gg \lambda_+$ and the
  physics will be dominated by forward scattering which can lead to a
  charge conducting but spin insulating phase.
  In a perturbative regime at higher temperatures we furthermore
  calculate the spin and charge densities around the domain wall and
  quantitatively discuss the interaction induced changes.
\end{abstract}


\maketitle

\section{Introduction}
In a one-dimensional electron system, particle-like excitations cannot
survive in the presence of interactions leading to a breakdown of
Fermi liquid theory. Instead, the excitations are of a collective,
bosonic nature and can be described by a universal low-energy
effective theory, the Luttinger liquid
(LL).\cite{PTP.5.544,luttinger:1154,Giamarchi} Impurities are expected
to play a particularly important role in one dimension, because
electrons are not able to circumvent them. Studies have indeed
revealed that impurities in many cases are relevant perturbations
effectively cutting the chain and therefore impeding transport at low
temperatures.\cite{KaneFisherPRL,PhysRevB.46.15233,PhysRevB.46.10866,PhysRevLett.75.934,PhysRevB.62.4370,SirkerLaflorencie,SirkerLaflorencie2}
More surprisingly, however, there are also situations where the
low-temperature behavior in the presence of multiple impurities
corresponds to ``healing'', i.e., to perfect
transmission.\cite{PhysRevB.46.10866,PhysRevB.46.15233}

One of the hallmarks of the Luttinger liquid is spin-charge
separation. This means that the normal modes of the Luttinger model
have either spin or charge character and are completely decoupled.
Spin-charge separation, however, only holds in the case of spin
degeneracy. In the presence of a magnetic field---which leads to spin
split bands---the normal modes of the Luttinger model acquire a mixed
character. This situation was first studied in
Refs.~\onlinecite{PhysRevB.47.6273,FrahmKorepin2} using the Hubbard
model in a magnetic field as a starting point. One obvious question to
ask is how impurities affect the low temperature properties in such a
system.  Since spin and charge are no longer decoupled we might expect
new low-energy fixed points which are not covered by the standard Kane
and Fisher picture.\cite{PhysRevB.46.15233}

Of experimental relevance is, in particular, the case of electrons in a quasi
one-dimensional wire coupled to bulk ferromagnetic order. Domain walls in the ferromagnet then act as
spatially extended magnetic impurities for the electrons in the wire.
Such systems of coupled electronic and magnetic degrees of freedom
have received considerable interest,\cite{Marrows2005} spurred, in
particular, by possible applications as magnetic domain-wall racetrack
memories.\cite{ParkinHayashi}
%
%
However, the focus has principally been on how the transport
properties of free electrons behave in a ferromagnetic wire with a
domain wall, and how these spin polarized currents set the domain wall
itself into motion. An interesting question to ask then is whether
electron-electron interactions can become important in such cases.
Aside from some work on mean-field\cite{PhysRevB.65.224419} and Hartree-Fock\cite{PhysRevB.74.224429,PhysRevB.76.205107} interactions, there is
little consideration in the literature of how the electronic and
magnetic behaviour of quasi one-dimensional ferromagnetic systems is
modified in a strongly correlated system.

In particular, the case of a ferromagnetic Luttinger liquid with a
domain wall present has not yet been fully addressed.  In the limit of
an infinitely sharp domain wall, Pereira and Miranda\cite{pereira04}
have considered an effective low temperature model containing only a
spin-flip back scattering term. Based on this model, they have argued
that the domain wall scattering in the ultimate low temperature limit
is the magnetic analog of the Kane-Fisher
problem.\cite{PhysRevB.47.6273,PhysRevB.46.15233,PhysRevB.46.10866}
In other words at low temperatures either a spin-charge insulator or a
Luttinger liquid is found.  In addition to the spin-flip back
scattering process considered in Ref.~[\onlinecite{pereira04}],
however, also a pure potential (spin-independent) back scattering term
is allowed by symmetry.\cite{PhysRevB.74.224429,PhysRevB.76.205107} By
starting from a model for an extended domain wall we will show that
such a term is indeed present and can be important for the physics at
very low temperatures. Our main focus, however, is what happens in the
more physically relevant regime of longer domain walls and higher
temperatures. The behaviour of the system in this case remains
unknown, and it is that question we wish to address here.

There are three possible temperature regimes in this problem and it is
convenient to introduce here some notation for them. At high
temperatures we have a ``perturbative regime'' where the domain wall
scattering can be treated as a small perturbation. As we consider
lower temperatures the perturbative treatment will break down due to
the presence of relevant operators. At first we may still consider the
domain wall as an extended region and we refer to this as the
``extended regime''. This regime is the focus of our study. At even
lower temperatures in the renormalization group flow the domain wall
will become effectively delta function like and we can treat the
relevant operators as boundary terms. This regime we refer to as the
``sharp regime'', and is the regime which has been previously
discussed in the
literature.\cite{pereira04,PhysRevB.74.224429,PhysRevB.76.205107} In
the sharp regime, the spin-flip and the potential back scattering
terms with scattering length $\lambda_+$ are the possible relevant
perturbations determining the low-temperature physics. We will show,
however, that if we start from a system with an extended domain wall
with length $\lambda\gg \lambda_+$, then temperatures $T/J <
\exp(-\lambda/\lambda_+)$, with $J$ being the magnetic exchange, are
required to enter this regime. This regime therefore is only
accessible if one already starts with a very sharp domain wall, a
situation which can possibly be realized by
nanoconstrictions.\cite{bruno}

Experimentally the construction of ferromagnetic chains of single
magnetic atoms is already possible.\cite{Gambardella} In such systems
the ferromagnetic order is observed to extend over small distances
separated by regions of
non-collinearity.\cite{Gambardella,PhysRevB.56.2340,PhysRevLett.73.898,PhysRevB.57.R677,RevModPhys.81.1495}
It is important to note that these chains are assembled on some
substrate and therefore cannot be considered as fully isolated
one-dimensional systems. Therefore the Mermin-Wagner
theorem,\cite{PhysRevLett.17.1133} which forbids long-range
ferromagnetic order at finite temperatures in a strictly
one-dimensional system with sufficiently short-range interactions,
does not apply. Furthermore, the substrate has consequences for the
effective spin exchange between the atoms. The spin exchange tensor
can quite generally be decomposed into a symmetric and an
antisymmetric part. In spin chains which are part of a regular
three-dimensional crystal the antisymmetric part is often forbidden by
inversion symmetry. For chains of single magnetic atoms on a
substrate, on the other hand, both terms are expected to be present so
that non-collinear spin order is a generic property of such systems.
The presence of the substrate might also disguise the Luttinger liquid
properties of the atomic wire and our model system might therefore be
too simplistic to directly apply to this situation.  Nevertheless, it
might serve as a starting point for the investigation of more
realistic models.

Other possible candidates our model might apply to include dilute
magnetic semiconductors,\cite{RevModPhys.78.809} and low temperature
ferromagnetic metals.\cite{metalwires,Granitzer2007302} Systems where the magnetic and electronic
degrees of freedom belong to different layers would also be a possible
realization. Furthermore, we want to point out that our analysis is
also valid for a quantum wire with a non-uniform external magnetic
field applied.

Our paper is organized as follows: In Sec.~\ref{mode} we introduce the
model and derive the low-energy effective theory by linearizing the
excitation spectrum followed by bosonization. In Sec.~\ref{lowt} we
consider the first order renormalization group (RG) equations for the
various scattering processes and discuss the fixed points of the RG
flow in the extended and sharp regimes. In Sec.~\ref{SpinCharge} we
study an experimentally accessible temperature regime where the
relevant scattering terms can be treated perturbatively. We discuss
different cases depending on the hierarchy of the different length
scales present in the problem and calculate the spin and charge
densities around the domain wall. In the final section we present a
brief summary of our results and some additional conclusions.

\section{The Model}
\label{mode}

We will consider an ``s-d'' like model in which the bulk magnetization
and the conduction electrons are treated separately (though of course
still coupled). We assume two timescales in the problem, a fast
electronic one and a slow magnetic one. This allows us to answer the
question of how the presence of the domain wall affects the Luttinger
liquid, forgetting the effect the motion of the domain wall will have
on the conduction electrons. For the already mentioned dilute magnetic
semiconductors and, in particular, ferromagnetic metals this model
with separate electronic and magnetic degrees of freedom, active on
different time scales, is a realistic starting point.

\begin{figure}
\includegraphics[width=0.45\textwidth]{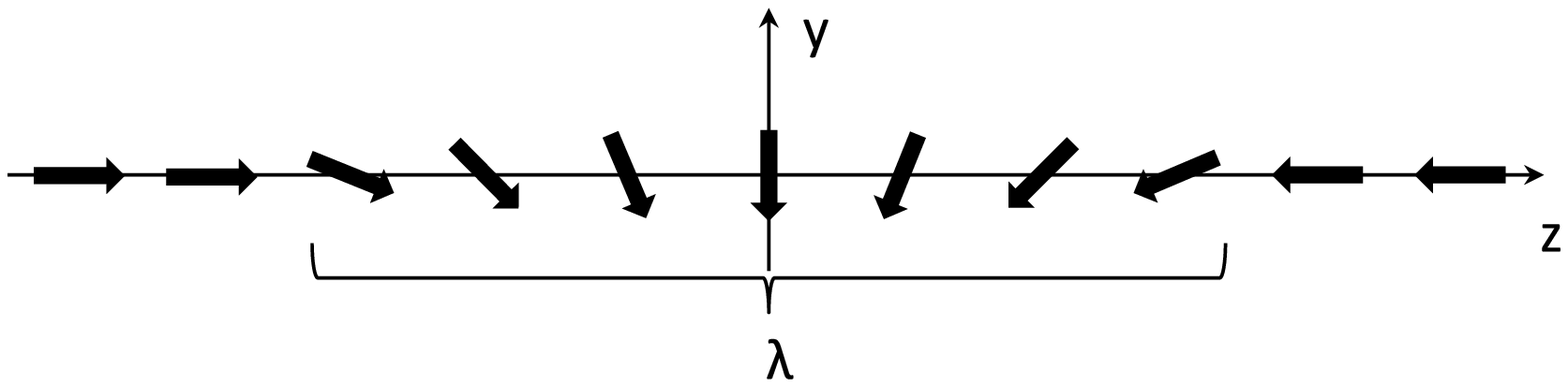}
\caption[Schematic]{A schematic view of the magnetization orientation in the wire. $\lambda$ is the length of the domain wall.}\label{schematic}
\end{figure}

The direction of the bulk magnetization of the wire can be described
by a unit vector
$\vec{n}(z)=\cos[\Theta(z)]\hat{\mathbf{z}}+\sin[\Theta(z)]\hat{\mathbf{y}}$
and we consider, in particular, the case
$\cos[\Theta(z)]=-\tanh[z/\lambda]$ describing the spatial profile of
a domain wall of length $\lambda$ situated at $z=0$, this is plotted
schematically in Fig.~\ref{schematic}. The magnetization is coupled
to the conduction band electrons with a strength given by the exchange
coupling $J$. We consider a screened, and hence short range,
interaction $V(z-z')$. To simplify the presentation we consider here a
spin-independent, $SU(2)$ symmetric interaction. We want to point out,
however, that the low-energy theory, Eqs.~(\ref{linHam}) and
(\ref{g-terms}), obtained after linearizing the excitation spectrum
remains valid for a spin-dependent interaction
$V_{\sigma\sigma'}(z-z')$ as long as the interaction is spin
conserving.

We start from the following standard ``s-d''
Hamiltonian,\cite{Blundell} $\tilde{H}=\tilde{H}_0+\tilde{H}_M+H_I$:
\begin{eqnarray}\label{fullhamiltonian}
\tilde{H}_0&=&\int\ud z\,\tilde{\psi}^\dagger_\sigma(z)\bigg[-\frac{1}{2m}\partial^2_{z}-\mu\bigg]\tilde{\psi}_\sigma(z),\\
\tilde{H}_M&=&-\frac{J}{2}\int\ud z\,\tilde{\psi}^\dagger_\sigma(z)\tilde{\psi}_{\sigma'}(z)\vec{\sigma}_{\sigma\sigma'}.\vec{n}(z),\nonumber\\
H_I&=&\frac{1}{2}\int\ud z\ud z'\tilde{\psi}^\dagger_\sigma(z)\tilde{\psi}^\dagger_{\sigma'}(z')V(z-z')\tilde{\psi}_{\sigma'}(z')\tilde{\psi}_\sigma(z).\nonumber
\end{eqnarray}
$\tilde{\psi}^\dagger_\sigma(z)$ is the creation operator for an
electron of spin $\sigma$ at a position $z$, $\mu$ is the chemical
potential, $m$ is the electron mass, and summation over the spin
indices $\{\sigma,\sigma'\}$ is implied. Due to the incommensurate
nature of the spin split Fermi points we gain no advantage from
explicitly considering half or full filling and the filling factor is
left general. We do exclude, however, the case of very small filling
where the Fermi energy $\varepsilon_F$ would become so small that a
linearization of the spectrum would only be appropriate at very low
temperatures. Furthermore, we only want to consider the case where the
filling in both bands is non-zero, i.e.~we are not interested in the
fully spin polarized ``half-metallic'' case\cite{halfmetallic} which would bring us back to an effective spinless
fermion model. In the following we set $\hbar=1$ and $k_B=1$.

\begin{figure}
\includegraphics*[width=0.45\textwidth]{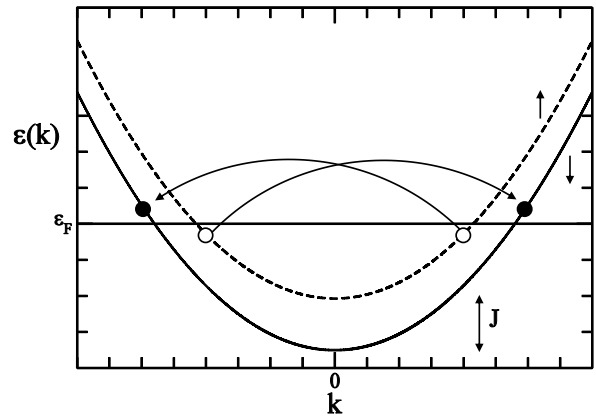}
\caption[Schematic Band Structure]{A schematic view of the spin split
  bands. One of the $g_{1\perp}$ processes is shown, which involve
  spin-flip scattering between left and right moving
  electrons.}\label{bandstructure}
\end{figure}

In order to be able to linearize the system our first step must be to
remove the spatially dependent, and in principle perhaps very large,
magnetization. This is achievable by rotating the spin direction
around the $x$-axis to a collinear ferromagnetic alignment via the
following gauge
transformation:\cite{PhysRevLett.78.3773,PhysRevB.16.4032}
$\mathbf{H}=\mathbf{U}^\dagger(z)\tilde{\mathbf{H}}\mathbf{U}(z)$,
$\psi_{\sigma}(z)=U^\dagger_{\sigma\sigma'}(z)\tilde{\psi}_{\sigma'}(z)$,
and $\mathbf{U}(z)=e^{\frac{i}{2}\Theta(z)\bm{\sigma}^x}$. The
interaction is left unaffected as it is $SU(2)$ invariant, the
magnetization is locally rotated to a Zeeman term $\tilde{H}_M\to
H_M$, and a gauge potential is introduced: $\tilde{H}_0\to H_0+H_G$.
Thus $H=H_0+H_G+H_I+H_M$ and
\begin{eqnarray}
  H_0&=&\sum_\sigma\int\ud z\,\psi^\dagger_\sigma(z)\bigg[-\frac{1}{2m}\partial^2_{z}-\mu\bigg]\psi_\sigma(z),\nonumber\\
  H_G&=&\frac{1}{8m}\sum_{\sigma\sigma'}\int\ud z\psi^\dagger_\sigma(z)[\Theta'(z)]^2\delta_{\sigma\sigma'}\psi_{\sigma'}(z)\nonumber\\&&
-\frac{1}{4m}\sum_{\sigma\sigma'}\int\ud z\bigg[\psi^\dagger_\sigma(z)i\Theta'(z)\sigma^x_{\sigma\sigma'}\partial_z\psi_{\sigma'}(z)\nonumber\\&&\qquad\qquad
-\big(\partial_z\psi^\dagger_\sigma(z)\big)i\Theta'(z)\sigma^x_{\sigma\sigma'}\psi_{\sigma'}(z)\bigg]
,\nonumber\\
  H_M&=&-\frac{J}{2}\sum_\sigma\int\ud z\,\psi^\dagger_\sigma(z)\sigma^z_{\sigma\sigma}\psi_\sigma(z) \label{hw}
\end{eqnarray}
with $\Theta'(z)=[\lambda\cosh(z/\lambda)]^{-1}$. The gauge
potential has been written in a manifestly Hermitian form.  The first
term of $H_G$ is a pure potential scatterer, the next two terms
describe spin-flip scattering. Without $H_G$ this is simply a spin
split band model,\cite{PhysRevB.47.6273} see Fig.~\ref{bandstructure}.
In our model the gauge potential introduces extended scattering terms,
active over the length of the domain wall, which have to be included
and are important for the low energy physics. The amplitudes of the
spin-flip scattering and the potential term are proportional to
$\lambda/\lambda_F$ and $(\lambda/\lambda_F)^2$, respectively, where
$\lambda_F$ is the Fermi wave length. We are here mainly interested in
the case $\lambda\gg \lambda_F$. Except for very low temperatures the
potential scattering term can then be safely neglected. On the other hand, in the limit
of a sharp domain wall $\lambda\to 0$, then
$\Theta'(z)\to\pi\delta(z)$, the scattering terms become boundary
operators, and both the spin-flip and the potential scatterer have to
be taken into account. We will discuss these issues in greater depth
in section \ref{lowt}.

The Hamiltonian (\ref{hw}) is now amenable to the usual bosonization
procedure,\cite{Giamarchi} the first step of which is linearization
via the ansatz $\psi_\sigma(z)=e^{ik_{F\sigma}z}\psi_{\sigma
  +}(z)+e^{-ik_{F\sigma}z}\psi_{\sigma -}(z)$, where
$k_{F\sigma}=\sqrt{2m(\mu\pm J/2)}$. The $+$ and $-$ indices denote
the right- and left-moving electrons, respectively.
Note that if
the Zeeman term is large we \emph{must} linearize around the spin
split Fermi points, leading to the breakdown of spin-charge
separation.
After linearization we have $H_0+H_M\to
H_{0\uparrow}+H_{0\downarrow}$, with the linearized Hamiltonians
\begin{eqnarray}
\label{linHam}
H_{0\sigma}= v_{F\sigma}\!\!\!\int\!\!\ud z[\psi^\dagger_{\sigma -}(z)i\partial_z\psi_{\sigma -}(z)-\psi^\dagger_{\sigma +}(z)i\partial_z\psi_{\sigma +}(z)]
\end{eqnarray}
where $mv_{F\sigma}=k_{F\sigma}$.

The interaction $H_I$ can be decomposed into spin parallel and spin
perpendicular components and written as
\begin{eqnarray}
\label{g-terms}
\tilde{H}_2&=&\sum_{\sigma, r=\pm}\int\ud z\bigg[\frac{\tilde{g}_{2\parallel\sigma}}{2}\rho_{\sigma r}\rho_{\sigma -r}+ \frac{g_{2\perp}}{2}\rho_{\sigma r}\rho_{\bar{\sigma} -r}\bigg],\\
H_4&=&\sum_{\sigma, r=\pm}\int\ud z\bigg[\frac{g_{4\parallel\sigma}}{2}\rho_{\sigma r}\rho_{\sigma r}+ \frac{g_{4\perp}}{2}\rho_{\sigma r}\rho_{\bar{\sigma} r}\bigg]\nonumber\textrm{, and}\\
H_1&=&\sum_{\sigma, r=\pm}\int\ud z\bigg[-\frac{g_{1\parallel\sigma}}{2}\rho_{\sigma r}\rho_{\sigma -r}\nonumber\\&&+ \frac{g_{1\perp\sigma}}{2}e^{2iz(k_{F\bar{\sigma}}-k_{F\sigma})} \psi^\dagger_{\sigma r}\psi^\dagger_{\bar{\sigma}-r}\psi_{\bar{\sigma}r}\psi_{\sigma -r}\bigg].\nonumber
\end{eqnarray}
Here we have suppressed the spatial indices and defined the local
density $\rho_{\sigma\pm}=\psi^\dagger_{\sigma\pm}\psi_{\sigma\pm}$.
Note that the ``g-ology'' given here refers to the already rotated
model, not the original physical picture. The chiral electrons of this
linearized model, physically speaking, have a non-collinear spin
orientation throughout the wire. Umklapp processes scattering two left
movers into right movers and vice versa are always neglected here due
to the non-commensurate nature of the Fermi wavevectors. We can
rescale the $g_{2\parallel\sigma}$ term to include the
$g_{1\parallel\sigma}$ process by redefining
$g_{2\parallel\sigma}=\tilde{g}_{2\parallel\sigma}-g_{1\parallel\sigma}$
with $\tilde{H}_2\to H_2$. The final $g_{1\perp\sigma}$ process,
schematically shown in Fig.~\ref{bandstructure}, can not be formulated
as a density-density interaction.

Finally we have our model to be bosonized. We introduce the chiral
bosonic fields $\phi_{\sigma r}(z)$.\cite{Giamarchi} The vertex
operator is
\begin{eqnarray}\label{boso}
\psi_{\sigma r}(z)=\frac{1}{\sqrt{2\pi\alpha}}e^{ir\sqrt{2\pi}\phi_{\sigma r}(z)}
\end{eqnarray}
where $\alpha$ is a short distance cutoff. This leads to the following
expression for the densities:
\begin{eqnarray}
\rho_{\sigma r}(z)=-\frac{1}{\sqrt{2\pi}}\partial_z\phi_{\sigma r}.
\end{eqnarray}
Thus we can write the quadratic part of the bosonic Hamiltonian,
$H_q$, which is composed from $H_0$, $H_M$, $H_2$, $H_4$ as a matrix
equation
\begin{eqnarray}
H_q&=&\int\ud z[\partial_z\Phi(z)]^T\mathbf{M}\partial_z\Phi(z).
\end{eqnarray}
$[\Phi(z)]^T=(\phi_{\uparrow+},\phi_{\uparrow-},\phi_{\downarrow+},\phi_{\downarrow-})$.
The bosonization procedure is thus sufficient to re-express all but
$H_{1\perp}$ and $H_G$ in terms of a diagonalizable quadratic bosonic
Hamiltonian. The matrix, $\mathbf{M}$, is
\begin{widetext}
\begin{eqnarray}
\mathbf{M}&=&\frac{1}{8\pi}\begin{pmatrix}
4\pi v_{F\uparrow}+2g_{4\parallel\uparrow} & g_{2\parallel\uparrow} & g_{4\perp} & g_{2\perp} \\
g_{2\parallel\uparrow} & 4\pi v_{F\uparrow}+2g_{4\parallel\uparrow} & g_{2\perp} & g_{4\perp} \\
g_{4\perp} & g_{2\perp} & 4\pi v_{F\downarrow}+2g_{4\parallel\downarrow} & g_{2\parallel\downarrow} \\
g_{2\perp} & g_{4\perp} & g_{2\parallel\downarrow} & 4\pi v_{F\downarrow}+2g_{4\parallel\downarrow}
\end{pmatrix}\nonumber.
\end{eqnarray}
\end{widetext}
$\mathbf{M}$ is a real symmetric matrix and as such has real
eigenvalues. We can now diagonalize $\mathbf{M}$ and will do so in several steps
to fully show the comparison with more standard expressions. In order
for our final normal modes to have positive velocities there is a
condition on this diagonalization procedure, given in the appendix.
However, for any realistic microscopic model this condition should
always be met, which can be shown explicitly in the case of the
Hubbard model in a magnetic field.\cite{PhysRevB.47.6273}

We can now make two unitary transformations. The first is
$\phi_{\sigma\pm}=(\phi_{\sigma}\mp \theta_{\sigma})/\sqrt{2}$. Note
that $\theta_{\sigma}$ is the adjoint of $\phi_{\sigma}$ and they
satisfy $[\phi_\sigma(z),\Pi_\sigma(z')]=i\delta(z-z')$ where
$\Pi_\sigma(z)=\partial_z\theta_\sigma(z)$. This first rotation has
the effect of uncoupling the two adjoint fields. The second
transformation is to rotate to the spin-charge representation:
$\phi_{c/s}(z)=[\phi_{\uparrow}(z)\pm\phi_{\downarrow}(z)]/\sqrt{2}$
(and similar for the $\theta(z)$ fields). The effect of these two
rotations can be summarized as
$\tilde{\mathbf{M}}=\tilde{\mathcal{U}}^{-1}\mathcal{U}^{-1}\mathbf{M}\mathcal{U}\tilde{\mathcal{U}}$
with $[\tilde{\Phi}(z)]^T=[\Phi(z)]^T\mathcal{U}\tilde{\mathcal{U}}$
(see equations \eqref{u} and \eqref{ut} in the appendix). If
$[\tilde{\Phi}(z)]^T=(\phi_{s},\phi_{c},\theta_{s},\theta_{c})$ then
our rotated Hamiltonian is defined by the matrix
\begin{eqnarray}
\label{M_SpinCharge}
\tilde{\mathbf{M}}&=&\frac{1}{2}\begin{pmatrix}
\frac{v_s}{K_s} & v_b  & 0 & 0 \\
v_b & \frac{v_c}{K_c} & 0 & 0 \\
0&0 & v_s K_s & v_a \\
0&0  & v_a & v_c K_c
\end{pmatrix}.
\end{eqnarray}
Here $K_s$ and $K_c$ are the spin and charge Luttinger parameters,
$v_s$ and $v_c$ are the spin and charge velocities, and $v_a$ and
$v_b$ describe the coupling between the spin and charge sectors. These
parameters are functions of the interaction strengths and Fermi
velocities, and to lowest order can be calculated directly, see
Eq.~\eqref{ks} in the appendix. At the non interacting $SU(2)$
symmetric point $K_s=K_c=1$. In the case of spin degeneracy we find,
as expected, $v_a=v_b=0$ and the spin and charge modes decouple.

The non-quadratic contributions from $H_G$ in this
representation are
\begin{eqnarray}\label{sch}
H^f_{Gsf}&=&\frac{g_f}{2\pi m\alpha}\int\ud z \Theta'(z)
\sin[\sqrt{2\pi}\theta_s(z)]\\&& \sin[z(k_{F\uparrow}-k_{F\downarrow})+\sqrt{2\pi}\phi_s(z)],\nonumber\\
H^b_{Gsf}&=&\frac{g_b}{2\pi m\alpha}\int\ud z\Theta'(z)
\sin[\sqrt{2\pi}\theta_s(z)]\nonumber\\&& \sin[-z(k_{F\uparrow}+k_{F\downarrow})+\sqrt{2\pi}\phi_c(z)],\nonumber\\
H^f_{Gp}&=&-\frac{1}{8m}\sqrt{\frac{2}{\pi}}\int\ud z [\Theta'(z)]^2
\partial_z\phi_c(z),\textrm{ and}\nonumber\\
H^b_{Gp}&=&-\frac{1}{8\pi m}\sum_\sigma\int\ud z g_{b\sigma}[\Theta'(z)]^2\nonumber\\&&
\cos[2zk_{F\sigma}+2\sqrt{\pi}\phi_\sigma(z)].\nonumber
\end{eqnarray}
The first term, $H^f_{Gsf}$, describes a forward scattering (upper
index ``f'') spin-flip (lower index ``sf'') process where a fermion is
exchanged between the spin up and spin down bands but stays on the
same side of the Fermi surface. The second term, $H^b_{Gsf}$, is a
backward scattering spin-flip term where a fermion is exchanged
between the bands and also moves from one side of the Fermi surface to
the other.  The third contribution $H^f_{Gp}$ is a potential (lower
index ``p''), spin conserving forward scattering process. 
The final term, $H^b_{Gp}$, is a spin conserving backward scattering
process. This is the scatterer considered by Kane and Fisher which
gives rise to the usual insulating fixed
point.\cite{PhysRevB.46.15233} All scattering processes are only
active over the length of the domain wall (or, more generally
speaking, the region of non-collinear spin order) where
$\Theta'(z)\neq 0$. The scattering coupling constants for forward and
backward spin-flip scattering from the domain wall are given by
$g_{f/b}= k_{F\uparrow}\pm k_{F\downarrow}$. We have also introduced
the bare potential scattering values $g_{b\sigma}=1/\alpha$ for
convenience.

The non-quadratic interaction term $H_{1\perp}$ is then
\begin{eqnarray}\label{sch2}
H_{1\perp}&=&\frac{2}{(2\pi\alpha)^2}\int\ud z\cos[2\sqrt{2\pi}\phi_s(z)]\\&&\big[g_{1\perp\uparrow}e^{-2iz(k_{F\uparrow}-k_{F\downarrow})}+g_{1\perp\downarrow}e^{2iz(k_{F\uparrow}-k_{F\downarrow})}\big].\nonumber
\end{eqnarray}
This last term also corresponds
to a backward scattering process. It stems, however, from the
interaction $H_I$ and therefore involves two fermions being scattered
between the different bands and from one side of the Fermi surface to
the other.  These contributions have their simplest interpretation in
terms of the spin and charge modes.  However, the spin and charge
modes are not eigenmodes of the model, see Eq.~(\ref{M_SpinCharge}),
and we also require Eqs.~\eqref{sch} and \eqref{sch2} in the appropriately rotated
basis.

The diagonalization of $\tilde{\mathbf{M}}$ introduces new velocities,
$u_i$, for $H_q$ and a set of $\{T_i^{\theta,\phi}\}$ parameters. It
can be summarized as
\begin{eqnarray}
\begin{pmatrix}\phi_c(z)\\ \phi_s(z)\end{pmatrix}&=&
\begin{pmatrix} \tilde{T}^{\phi}_{1} & \tilde{T}^{\phi}_{2} \\ T^{\phi}_{1} & T^{\phi}_{2} \end{pmatrix}
\begin{pmatrix}\phi_1(z)\\ \phi_2(z)\end{pmatrix}\nonumber
\textrm{ and}\\
\begin{pmatrix}\theta_c(z)\\ \theta_s(z)\end{pmatrix}&=&
\begin{pmatrix} \tilde{T}^{\theta}_{1} & \tilde{T}^{\theta}_{2} \\ T^{\theta}_{1} & T^{\theta}_{2} \end{pmatrix}
\begin{pmatrix}\theta_1(z)\\ \theta_2(z)\end{pmatrix}
\end{eqnarray}
with the parameters $\{T_i^{\theta,\phi}\}$ as given in
Eq.~\eqref{rotate1} of the appendix. They are all known in terms of
the previously mentioned Luttinger parameters. We have also
simultaneously rescaled the fields to obtain two distinct eigenvalues
rather than four.  The final Hamiltonian is
$H=H_q+H_{1\perp}+H^f_{Gsf}+H^b_{Gsf}+H^b_{Gp}+H^f_{Gp}$ where
\begin{eqnarray}
H_q&=&\frac{u_i}{2}\int\ud z\,\left[(\partial_z\phi_i(z))^2+(\Pi_i(z))^2\right]
\end{eqnarray}
and, with summation over $i$ and $j$ implied:
\begin{eqnarray}
H^f_{Gsf}&=&\frac{g_f}{2\pi m\alpha}\int\ud z\, \Theta'(z)
\sin[\sqrt{2\pi}T_i^\theta\theta_i(z)]\\&& \sin[z(k_{F\uparrow}-k_{F\downarrow})+\sqrt{2\pi}T_j^\phi\phi_j(z)],\nonumber\\
H^b_{Gsf}&=&\frac{g_b}{2\pi m\alpha}\int\ud z\,\Theta'(z)
\sin[\sqrt{2\pi}T_i^\theta\theta_i(z)]\nonumber\\&& \sin[-z(k_{F\uparrow}+k_{F\downarrow})+\sqrt{2\pi}\tilde{T}_j^\phi\phi_j(z)],\nonumber\\
H^f_{Gp}&=&-\frac{1}{8m}\sqrt{\frac{2}{\pi}}\int\ud z [\Theta'(z)]^2
\tilde{T}^\phi_i\partial_z\phi_i(z),\textrm{ and}\nonumber\\
H^b_{Gp}&=&-\frac{1}{8\pi m}\sum_\sigma\int\ud z g_{b\sigma}[\Theta'(z)]^2\nonumber\\&&
\cos[2zk_{F\sigma}+\sqrt{2\pi}(\tilde{T}^\phi_i+\sigma_{\sigma\sigma}^zT^\phi_i)\phi_i(z)]\nonumber
\end{eqnarray}
for the scattering terms and finally:
\begin{eqnarray}
H_{1\perp}&=&\frac{2}{(2\pi\alpha)^2}\int\ud z\cos[2\sqrt{2\pi}T_i^\phi\phi_i(z)]\nonumber\\&&\big[g_{1\perp\uparrow}
e^{-2iz(k_{F\uparrow}-k_{F\downarrow})}+g_{1\perp\downarrow}e^{2iz(k_{F\uparrow}-k_{F\downarrow})}\big].\nonumber
\end{eqnarray}
The appropriate excitations of such an $SU(2)$ {\it asymmetric} model
have no obvious physical interpretation. This effective bosonic field
theory lays the foundation of our further analysis.  A similar model
is found by Braunecker et
al.\cite{PhysRevLett.102.116403,PhysRevB.80.165119} in a different
situation, and of course by Pereira and Miranda\cite{pereira04} but
without the $H^f_{Gsf}$ and $H_{1\perp}$ terms which do not play any
role for very sharp domain walls where the length scale $\lambda$ is
no longer present. However they also neglect $H^b_{Gp}$ which does
play a role in the sharp wall limit. Indeed it is this term which,
when dominant, leads to the Kane and Fisher insulating fixed point.
This will become clear in Sec.~\ref{lowt} where the case of a sharp
domain wall is obtained as a specific limit in our general analysis.

\section{Low Energy Physics}
\label{lowt}
In the generic case, we have three natural length scales present in
the problem: $\lambda_+\sim (k_{F\uparrow}+k_{F\downarrow})^{-1}$
related to spin-flip backward scattering, $\lambda_-\sim
(k_{F\uparrow}-k_{F\downarrow})^{-1}$ related to spin-flip forward
scattering, and the domain wall length $\lambda$. In the limit of weak
magnetization $\lambda_-\to\infty$ and $2\lambda_+\to\lambda_F$ while,
on the other hand, the limit of large magnetization, when one spin
channel becomes frozen out, gives
$\lambda_+\approx\lambda_-\approx\lambda_F$.  It is crucial for the
further analysis to observe that the relative importance of the
forward and backward scattering terms is now not only determined by
their scaling dimensions but also by the hierarchy of the three
different length scales. Furthermore, in the RG flow we must
distinguish between the extended and sharp regimes. In the beginning
of the RG flow we have an extended domain wall and the scattering
terms can be treated as bulk terms of dimension $d=2$.  However, when
the ultraviolet momentum cutoff $\Lambda$ during the RG process becomes of the
order $1/\lambda$ then the extended domain wall will begin to look
effectively point like, i.e.~of dimension $d=1$. The scattering terms
then become boundary operators. At this stage the direction and rate
of the flow of all of the operators can change, leading to the final
low-temperature fixed points. For this effective $d=1$ flow we must
take the result of the $d=2$ flow as the ``zeroth order'' coupling
constants. In physical terms, the domain wall starts to look
point-like if the electrons are correlated spatially over lengths much
larger than the domain wall length, i.e., if $J/T\gg\lambda/\alpha$.

Another important point to note is that for an extended domain wall
we usually have $\lambda\gg\lambda_+$, i.e., the backward scattering
terms are strongly oscillating over the length of the domain wall.
This leads to very small bare effective backward scattering couplings
which can be estimated as follows: The derivative of the domain wall
profile, $\Theta'(z)$, can be Fourier transformed leading to
$\Theta'(k)=\pi/\cosh(\pi k\lambda/2)$. The non-oscillating
component of the backward scattering amplitude is then proportional to
$g_b\Theta'(k=\pm 1/\lambda_+)$. Since $\Theta'(k)$ is a function
which is sharply peaked at $k=0$ for long domain walls, we have
$|g_b\Theta'(k=\pm 1/\lambda_+)|\sim
|g_b\exp(-\lambda/\lambda_+)|\ll 1$. If backward scattering is
relevant, then the effective coupling constant will grow under the RG
flow as
\begin{equation}
g_b^{\rm eff} \sim g_b\exp(-\lambda/\lambda_+)\left(\frac{T_0}{T}\right)^{d-\gamma_b}
\label{backward_scale}
\end{equation}
where $T_0$ is an energy scale of order $J$. Backward scattering will
only have an appreciable effect if the initially small bare coupling
has again become of order $1$ under the RG flow. This requires
temperatures $T/J < \exp(-\lambda/\lambda_+)$ which are extremely
small for many realistic situations. We therefore expect that forward
scattering---ignored in previous investigations of the domain wall
problem---will play the dominant role in these cases.

Before discussing the various regimes any further, we want to derive
the first order RG equations for the forward and backward scattering
terms. We start by writing a functional integral partition
function\cite{NegeleOrland}
\begin{equation}
\mathcal{Z}=\int D\phi_i D\Pi_i e^{-\int_0^{ \beta} \ud \tau\big[\int\ud z[-i\Pi_i(z)\partial_\tau\phi_i(z)]+H[\Pi_i(z),\phi_i(z)]\big]}
\end{equation}
with periodic boundary conditions in imaginary time $\tau$. Following
the standard procedure we split the fields into fast, $\phi^>$, and
slow, $\phi^<$, fields. Our fast fields are defined for
$\Lambda'<|k|,|\omega|/u_i<\Lambda$, and the slow for
$|k|,|\omega|/u_i<\Lambda'$, with $\Lambda$ an ultraviolet cut-off.
Expanding the exponent in terms of $H_{1\perp}$, $H^{f,b}_{Gsf}$ and $H^{b}_{Gp}$ and performing the
averaging over the fast modes we then re-exponentiate the expression
to find the appropriate scaling equations.  The flow is parametrized in terms of $l$, defined as
$\Lambda=\Lambda_0e^{-l}$ and $\Lambda'=\Lambda_0e^{-l-\delta l}$.

We find for the $H_{1\perp}$ term to first order that
\begin{eqnarray}
\frac{\ud g_{1\perp\sigma}}{\ud l}=
g_{1\perp\sigma}\big[2-\underbrace{2[(T^\phi_1)^2+(T^\phi_2)^2]}_{\equiv \gamma_1}\big].
\end{eqnarray}
This term is an irrelevant perturbation ($\gamma_1>2$) for any
realistic situation we are here interested in. In the limit of weak
magnetization we can simplify the expression to find $\gamma_1=2K_s$.
In this limit $K_s>1$ and it becomes clear that the term is
irrelevant.

The same analysis is performed on the domain wall scattering terms.
For spin-flip back scattering we find:
\begin{eqnarray}
\label{dgb}
\frac{\ud g_b}{\ud l}=
g_b\big[d-\underbrace{\frac{1}{2}[(\tilde{T}^\phi_1)^2+(\tilde{T}^\phi_2)^2+(T^\theta_1)^2+(T^\theta_2)^2]}_{\equiv
\gamma_b}\big].
\end{eqnarray}
As already mentioned $d$ is the dimension of the wall, for spatially
extended walls this is $2$, whereas it is $1$ in the limit of an
infinitely sharp wall. As usual the perturbation is relevant
(irrelevant) if $\gamma_b <d$ ($\gamma_b>d$). In the limit of $J\to 0$ we can simplify this expression to $\gamma_b =
[(K_s)^{-1}+K_c]/2$ consistent with Refs.~\onlinecite{pereira04},
\onlinecite{pereirathesis}. If we consider the Luttinger parameters
for the Hubbard model in the zero magnetization limit we see that this
term is always relevant for repulsive interactions both for $d=1$ and
$d=2$ but can also become irrelevant in both cases for attractive
interactions. In the general case described by Eq.~\eqref{dgb} the
relevance or irrelevance of backward scattering depends on the set of
rotation parameters $T_i^{\theta,\phi}$ and no general conclusions are
possible.

Similarly the spin-flip forward scattering equation stands as
\begin{eqnarray}
\label{dgf}
\frac{\ud g_f}{\ud l}=
g_f\big[d-\underbrace{\frac{1}{2}[(T^\phi_1)^2+(T^\phi_2)^2+(T^\theta_1)^2+(T^\theta_2)^2]}_{\equiv
\gamma_f}\big].
\end{eqnarray}
In the limit of $J\to 0$ this simplifies to $\gamma_f =
[(K_s)^{-1}+K_s]/2$. Since $K_s\geq 1$, in this limit it follows that
$\gamma_f\geq 1$. For the case of a sharp wall ($d=1$) forward
scattering is therefore always
irrelevant.\cite{pereira04,pereirathesis} In the case of an extended
wall ($d=2$), on the other hand, forward scattering is relevant in
this limit if $K_s<2+\sqrt{3}$. The generic case described by
Eq.~\eqref{dgf} is again very complicated to analyze. However, at
least for simple microscopic models and in the limit of weak
interactions where the rotation parameters can be calculated
explicitly (see Eqs.~\eqref{A6}, \eqref{A7} in the appendix), we find
that forward scattering is always relevant for $d=2$.

The potential back scattering equations are, for the two spin channels,
\begin{eqnarray}
\label{dgbp}
\frac{\ud g_{b\sigma}}{\ud l}=
g_{b\sigma}\big[d-\underbrace{\frac{1}{2}[(\tilde{T}^\phi_1\pm T^\phi_1)^2+(\tilde{T}^\phi_2\pm T^\phi_2)^2]}_{\equiv
\gamma_{b\sigma}}\big]
\end{eqnarray}
with the plus (minus) sign applying for $\sigma=\uparrow$
($\sigma=\downarrow$). In the limit of $J\to 0$ we find
$\gamma_{b\sigma} =[K_c+K_s]/2$. In the isotropic limit, $K_s\to 1$,
this term is always relevant for repulsive interactions. In general
however it can be either relevant or irrelevant. Indeed, it is also
possible that potential backward scattering is relevant for one spin
channel and irrelevant for the other.

The potential forward scattering term has scaling dimension $x=1$. It
will therefore be relevant for an extended domain wall and marginal in
the limit of a sharp wall.

\begin{figure}
\includegraphics*[width=0.45\textwidth]{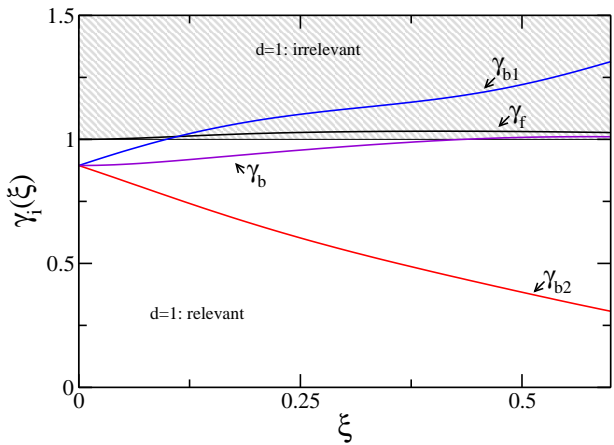}
\caption[$\gamma_{f,b}$]{(Color online) $\gamma_{f,b}$ and
  $\gamma_{b\sigma}$ for forward and backward, spin flip and
  potential, scattering from an extended domain wall obtained for a
  specific $U$-$V$ model (see text for details) as a function of
  $\xi=J/2\varepsilon_F$. The regions where the operators are relevant
  or irrelevant for $d=1$ are marked. The condition $K_s=1$ has been
  imposed by hand at the $SU(2)$ symmetric point ($J=0$), see
  appendix.}\label{gammas}
\end{figure}

As an example, we present in Fig.~\ref{gammas} the two scaling
dimensions for forward and backward spin-flip scattering off an
extended wall as a function of $\xi=J/2\varepsilon_F$ calculated for
a $U$-$V$ model with an on-site interaction $U/t=1.10$, and nearest
neighbor interaction $V/t=0.11$, where $t$ is the hopping amplitude.
Here the parameters of our low-energy effective model are determined
in a lowest order expansion in the interaction parameters.
We want to remind the reader that by rotating back to the original
physical model one finds that ``spin-flip forward scattering'' refers
to an electron which passes through the wall \emph{without} changing
its spin.

\subsection{RG flow in the extended domain wall regime}
At first we will focus on the $d=2$ case. For extended domain walls
($\lambda\gg\lambda_F$) the bare potential scattering terms $\sim
(\lambda/\lambda_F)^2$ are much smaller than the spin-flip scattering
terms $\sim(\lambda/\lambda_F)$. In the temperature range where it is
appropriate to use the RG flow with $d=2$ we can therefore neglect the
potential scattering terms. They will, however, become important for
the RG flow with $d=1$ at even lower temperatures discussed in the
next subsection. 

Though with the physical parameters we consider in section
\ref{SpinCharge} we find both $g_{f,b}$ to be relevant, by varying the
$v_{c,s}$, $K_{c,s}$, and $v_{a,b}$ parameters we can find regimes
where forward scattering remains relevant but backscattering becomes
irrelevant.

We can now identify several regimes. We focus here first on the small
magnetization limit close to half-filling where
$\lambda_-\gg\lambda_+$. For the backward scattering Hamiltonian,
$H_{Gsf}^{b}$, Eq.~\eqref{sch}, this means that we take the limit
where $k_{F\uparrow}\approx k_{F\downarrow}\to k_F\approx \pi/2$. We
can then simplify \eqref{sch} and find
\begin{eqnarray}
H_{Gsf}^{b}&=&\frac{g_b}{2\pi m\alpha}\int\ud z\Theta'(z)\sin[\sqrt{2\pi}\theta_s(z)]\nonumber\\&& \cos[2zk_F]\sin[\sqrt{2\pi}\phi_c(z)].
\end{eqnarray}
We now consider the case that this term is relevant, i.e., $g_b$
becomes large at low temperatures. Firstly, for `narrow enough'
walls, i.e., for domain walls of the order of the Fermi wave length
$\lambda_F$ when the $2k_F$ oscillations do not cancel out the
contributions, then in order to minimize the energy the fields become
locked over the length of the domain wall in the values
$\{\phi_c=\sqrt{2\pi}(m+\frac{3}{4}),\theta_s=\sqrt{2\pi}(n+\frac{1}{4})\}$
or
$\{\phi_c=\sqrt{2\pi}(m+\frac{1}{4}),\theta_s=\sqrt{2\pi}(n+\frac{3}{4})\}$
for integer $m,n$.
Therefore the domain wall becomes an impenetrable barrier for both
charge and spin excitations and we find a spin charge
insulator.\cite{pereira04,PhysRevB.51.17827}
If we consider longer domain walls then, as already discussed
previously, the effective bare coupling for backward scattering will
be close to zero. Hence, for extended domain walls, $\lambda \gg
\lambda_F$, forward scattering is always the more important because
$\lambda_+\ll \lambda_-$.

Therefore there is also a regime in which only forward scattering is
relevant: either the long domain wall case or the case of irrelevant
backward scattering ($\gamma_b>2$).  Considering again the limit of
small magnetization and a system close to half-filling, the forward
scattering term in Eq.~\eqref{sch} simplifies to
\begin{eqnarray}
H_{Gsf}^f&=&\frac{g_f}{2\pi m\alpha}\int\ud z\Theta'(z)\sin[\sqrt{2\pi}\theta_s(z)]\nonumber\\&& \cos[z(k_{F\uparrow}-k_{F\downarrow})]\sin[\sqrt{2\pi}\phi_s(z)].
\end{eqnarray}
In order to minimize the energy in this case the fields become locked
over the length of the domain wall in the values
$\{\phi_s=\sqrt{2\pi}(m+\frac{3}{4}),\theta_s=\sqrt{2\pi}(n+\frac{1}{4})\}$
or
$\{\phi_s=\sqrt{2\pi}(m+\frac{1}{4}),\theta_s=\sqrt{2\pi}(n+\frac{3}{4})\}$
for integer $m,n$. In this scenario only the spin sector is frozen out
and we find a C$1$S$0$ phase where the spin mode is gapped but charge
excitations remain gapless.\cite{PhysRevB.53.12133} In the physical
reference frame this is a state in which the incoming spin current
does not scatter from the domain wall and, after traversing the domain
wall, ends in the anti-parallel spin configuration with respect to the
bulk. This means that the domain wall profile can no longer be taken
as adiabatic.

From the above considerations we have several possibilities. Firstly
we can have backward scattering as either relevant or irrelevant.
Secondly we must consider the relative length scales. If
$\lambda_+<\lambda$ then the backward scattering terms are small due
to averaging over their oscillations. A similar case holds for the
forward scattering with $\lambda_+\to\lambda_-$ in the preceding. For
the case in which both the forward and backward scattering
length-scales are shorter than the domain wall length we end up in the
completely adiabatic limit, as one would expect, and the system shows
Luttinger liquid properties (``adiabatic LL''). Again we note that at
extremely low temperatures we have to switch to a $d=1$ RG flow and
backward scattering processes will begin to dominate and can lead to
insulating regimes. In principle, we can also be in the opposite
regime when both forward and backward scattering length-scales are
longer than the domain wall length in which case the scaling
dimensions of the forward and backward scattering terms alone
determine what the low-energy fixed point is.  Note that this is
possible without requiring a very sharp delta function like domain wall profile.
In general, however, the low-temperature phase the system ends up in in the
extended regime depends not only on the relevance of the operators,
but also on the hierarchy of length scales.

 The different behaviour in the extended regime which could be
 identified from the first order RG equations are summarized in table
 \ref{tab1}.
\begin{table}
  \caption{\label{tab1} The different phases of the ferromagnetic Luttinger liquid with an extended domain wall ($d=2$, relevant spin-flip forward scattering) depending on the scaling dimensions of the spin-flip
    backward scattering operator {\it and} the hierarchy of the three length scales present in the problem.}
\begin{ruledtabular}
{ \renewcommand{\arraystretch}{1.5}
 \renewcommand{\tabcolsep}{0.2cm}
\begin{tabular}{c|c|c|c}
& $\lambda\lesssim \lambda_\pm$&$\lambda_+<\lambda<\lambda_-$&$\lambda_\pm<\lambda$\\
\hline
$2-\gamma_b >0$ & Spin and charge & C$1$S$0$ & Adiabatic \\
  & insulator &  & LL \\\hline
$2-\gamma_b <0$  & C$1$S$0$ & C$1$S$0$ & Adiabatic \\
&&& LL
\end{tabular}}
\end{ruledtabular}
\end{table}
Finally, let us also comment on the case of a generic magnetization
and arbitrary filling. In this case the analysis above stays valid,
the spin and charge modes, however, get locked into more complicated
spin and charge density wave states over the length of the domain
wall.

\subsection{Fixed points of the $d=1$ sharp domain wall regime}

For the case of a very sharp domain wall ($d=1$) we have shown that
forward spin-flip scattering is always irrelevant. The forward
potential scattering term is marginal and we will ignore it as well.
In this case the possible phases are therefore determined by the
scaling dimensions of the two backward scattering terms alone.
(Naturally for a sharp wall the length-scales can play no further
role.) Formally one can find the sharp domain wall limit from
Eq.~(\ref{hw}) by taking the limit $\lambda\to 0$, which requires
$\Theta'(z)\to\pi\delta(z)$. This leads to an effective model
where all the boundary scattering terms allowed by symmetry are present. A full
description of the phase space for this model with different relevant
perturbations present requires the solution to the second order RG
equations to find the separatrix between the different low temperature
fixed points.  The second order equation for our model is more
complicated than for the standard sine-Gordon model. A diagonal
equation in the $\phi_i$'s is not recovered and to perform any further
analysis we would have to re-diagonalize the problem and then
renormalize the model once again, repeating these steps until we reached the fixed
point.\cite{PhysRevB.81.165402} This is perhaps not totally unexpected
as the scattering terms we are dealing with explicitly couple the
normal modes. We leave the more involved second order RG analysis to a
future work and focus here on what the first order equations can tell
us. 
The flow of a similar model has already been analyzed by Ara\'ujo, et
al.\cite{PhysRevB.74.224429,PhysRevB.76.205107} using poor man's
scaling. In that work they consider both spin-flip and pure potential
backscattering from a sharp domain wall, treating the interaction only
perturbatively. They find phases dominated by the spin-flip and pure
potential backscattering processes.  In contrast Pereira and
Miranda\cite{pereira04} consider only the spin-flip backscattering
term and hence cannot recover all possible low energy phases.


There are three possible fixed points of the RG flow depending on the
relative relevance of the three backscattering channels, $g_b$,
$g_{b\uparrow}$, and $g_{b\downarrow}$. The system can flow to a spin
and charge insulator, an effectively spinless Luttinger liquid, or a
spin-full Luttinger liquid. Furthermore the spin and charge insulator
itself can show different physical behaviour in the region of the
domain wall depending on the relative relevance of the backscattering
operators. This behaviour will be confined to some region around the
boundary. Firstly let us discuss the spin and charge insulating phase.
If the pure potential backscattering for both spin channels is
relevant and dominates, $\gamma_{b\sigma}< \gamma_{b}$, then the
system will flow to the spin and charge insulating fixed point already
studied by Kane and Fisher\cite{PhysRevB.46.15233} with the spin part
of the system remaining completely unaffected. If the spin-flip back
scattering term is relevant and dominates, $\gamma_{b}<
\gamma_{b\sigma}$, then we have the fixed point analyzed by Pereira
and Miranda.\cite{pereira04} The spin-flip back scattering will tend
to equilibriate the number of up and down spins. This fixed point must
therefore correspond physically to a spin and charge insulator with a
region of reduced spin polarization around the boundary. It is also
possible to have the situation where $\gamma_{b\sigma}< \gamma_{b}<
\gamma_{b\bar\sigma}$ with $\gamma_{b\sigma},\gamma_b<1$. In this case
as the system approaches the fixed point first one spin channel will
become insulating, however, before the second spin channel also
becomes insulating the spin-flip backscattering term will tend to
align the spins into the first channel. Once the spins are scattered
into the first channel they are more likely to be scattered without a
spin flip than back into the other spin channel. Therefore the final
fixed point will likely be an insulator with a region of increased
spin polarization around the domain wall.  Secondly if potential back
scattering is relevant for one channel but irrelevant for the other
and spin-flip scattering is irrelevant as well, then we are left with
an effective spinless Luttinger liquid, i.e., we have one insulating
and one conducting channel. Finally, if all back scattering operators
are irrelevant, then the system remains a Luttinger liquid.

These results are summarized in table \ref{tab2}.
\begin{table}
  \caption{\label{tab2} The different phases of the ferromagnetic Luttinger liquid with an effectively sharp domain wall ($d=1$, irrelevant forward scattering) depending on the relevance or irrelevance of the 
    backward scattering operators. The physics of the system surrounding the domain wall for the spin and charge insulating phase depends upon the relative relevance of the backward scattering operators and is discussed fully in the text.}
\begin{ruledtabular}
{ \renewcommand{\arraystretch}{1.5}
 \renewcommand{\tabcolsep}{0.2cm}
\begin{tabular}{c|c|c}
 $\gamma_{b}<1$ & $\gamma_{b\sigma}<1$ &  $\gamma_{b}>1$\\
and/or $\gamma_{b\uparrow},\gamma_{b\downarrow}<1$ & $\gamma_b,\gamma_{b\bar{\sigma}}>1$ &  $\gamma_{b\uparrow},\gamma_{b\downarrow}>1$\\
\hline
Spin and charge &  &   \\
insulator & Effective spinless LL  & LL  \\
\end{tabular}}
\end{ruledtabular}
\end{table}

\section{Spin and Charge Density in the perturbative regime}
\label{SpinCharge}
We now return to the case of an extended domain wall,
$\lambda\gg\lambda_F$, where the potential scattering terms can be
ignored and consider a perturbative temperature regime where the spin
flip scattering terms can be treated perturbatively. This allows us to
calculate the spin and charge densities around the domain wall. Spin
and charge density oscillations around impurities are not only
experimentally relevant, but also provide a useful theoretical tool to
analyze the dominant physical scattering processes in general low
dimensional strongly correlated
systems.\cite{EggertRommer,AnfusoEggert,EggertSyljuasen,SirkerLaflorencieEPL}
As we are interested in the case where both forward and backward
scattering are relevant perturbations, a perturbative treatment of the
scattering terms will break down at low enough temperatures. We indeed
find that the perturbative corrections increase as a power law in
inverse temperature. In order for perturbation theory to be valid, we
find that the following conditions, for forward and backward
scattering terms respectively, have to be held:
\begin{eqnarray}\label{famp}
\left(\frac{T}{T^*_1}\right)^{\alpha_1}\left(\frac{T}{T^*_2}\right)^{\alpha_2} & \ll& \frac{4\pi \lambda mT}{g_f[T^\phi_1T^\theta_1+T^\phi_2T^\theta_2]}
\end{eqnarray}
and
\begin{eqnarray}\label{bamp}
\left(\frac{T}{T^*_1}\right)^{\beta_1}\left(\frac{T}{T^*_2}\right)^{\beta_2} & \ll& \frac{4\pi \lambda mT}{g_b[T^\phi_1T^\theta_1+T^\phi_2T^\theta_2]}.
\end{eqnarray}
Here $T^*_{i}\sim u_i/(2\pi\alpha)$ is a cutoff scale and the
exponents are given by $\alpha_i=[(T^\phi_i)^2+(T^\theta_i)^2]/4$ and
$\beta_i=[(\tilde{T}^\phi_2)^2+(T^\theta_2)^2]/4$. Note that
$\alpha_1+\alpha_2 =\gamma_f/2$ and $\beta_1+\beta_2=\gamma_b/2$ so
that the scaling dimensions show up in Eqs.~\eqref{famp} and
\eqref{bamp} in the expected way.


We want to present our results in the physical unrotated frame. The
perturbative results are, however, obtained in the rotated frame so
that we have to use the gauge transformation $\mathbf{U}(z)$ once
more. The spin density, in the physical frame, is
\begin{eqnarray}
  \vec{S}(z)=\frac{1}{2}{\tilde{\bm\psi}}^\dagger(z)\vec{{\bm\sigma}}{\tilde{\bm\psi}}(z)  = \frac{1}{2}{\bm\psi}^\dagger(z) \underbrace{{\bm U}^\dagger\vec{{\bm\sigma}}{\bm U}}_{\vec{{\bm\sigma}}'}{\bm\psi}(z)
\end{eqnarray}
where the gauge rotated Pauli matrices are given explicitly by
\begin{eqnarray}
{\bm\sigma}'_x&=&{\bm\sigma}^x,\nonumber\\
{\bm\sigma}'_y&=&\cos[\Theta(z)]{\bm\sigma}^y -\sin[\Theta(z)]{\bm\sigma}^z,
\nonumber\\
{\bm\sigma}'_z&=&\sin[\Theta(z)]{\bm\sigma}^y+\cos[\Theta(z)]{\bm\sigma}^z.
\end{eqnarray}
Using these relations, the spin densities in the physical frame
$\vec{S}(z)$ can now be constructed from the spin densities
$\vec{S}_0(z)=\frac{1}{2}{\bm\psi}^\dagger(z)\vec{{\bm\sigma}}{\bm\psi}(z)$
in the rotated frame.
The corrections to the bulk in first order in forward and backward
scattering off the domain wall are given by
\begin{eqnarray}\label{spin0x}
\langle \Delta S^x_0(z)\rangle&=&\frac{1}{\alpha\pi}\big\langle\cos[\sqrt{2\pi}\theta_s(z)]\times\\&&
\big[\cos[(k_{F\uparrow}-k_{F\downarrow})z+\sqrt{2\pi}\phi_s(z)] \nonumber\\&&+\cos[(k_{F\uparrow}+k_{F\downarrow})z+\sqrt{2\pi}\phi_c(z)]\big]\big\rangle,\nonumber\\ \label{spin0y}
\langle \Delta S^y_0(z)\rangle&=&\frac{1}{\alpha\pi}\big\langle\sin[\sqrt{2\pi}\theta_s(z)]\times\\&&
\big[\cos[(k_{F\uparrow}-k_{F\downarrow})z+\sqrt{2\pi}\phi_s(z)] \nonumber\\&&+\cos[(k_{F\uparrow}+k_{F\downarrow})z+\sqrt{2\pi}\phi_c(z)]\big]\big\rangle\textrm{, and\nonumber}\\ \label{spin0z}
\langle \Delta S^z_0(z)\rangle&=&\bigg\langle-\frac{1}{\sqrt{2\pi}}\partial_z\phi_s(z)\\&&+\frac{1}{2\alpha\pi}\sum_{\sigma}\sigma^z_{\sigma\sigma} \cos[2xk_{F\sigma}-2\sqrt{\pi}\phi_\sigma(z)] \bigg\rangle.\nonumber
\end{eqnarray}
Note that in the linearization procedure, strictly speaking, we should
write $\psi_\sigma(z)=\psi_{0\sigma}(z)+e^{ik_{F\sigma}z}\psi_{\sigma
  +}(z)+e^{-ik_{F\sigma}z}\psi_{-\sigma}(z)$. As such there are
${\bm\psi}_0^\dagger(z)\vec{{\bm\sigma}}{\bm\psi}_0(z)$ terms missing
from the above spin densities which give the bulk values. We can also
use a description where we absorb the effective magnetic field by a
shift in the bosonic fields instead of linearizing around the spin
split Fermi points.  In this case, however, we must take into account
curvature terms.\cite{1742-5468-2007-08-P08022} This would reintroduce
the bulk spin density terms explicitly in Eqs.~\eqref{spin0x} to
\eqref{spin0z}.

\begin{figure}
\includegraphics*[width=0.45\textwidth]{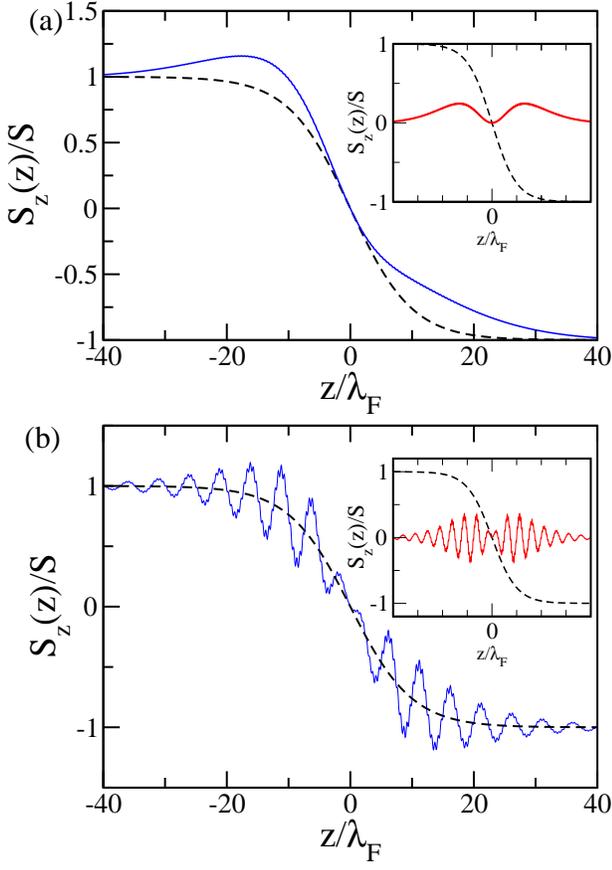}
\caption[Spin Density]{(Color online) The spin density $S_{z}(z)/S$,
  showing the zeroth order (dashed line) term and the total value
  (blue, solid line). The inset shows a comparison of the the zeroth order
  (dashed line) and first order (red, solid line) terms. Zeroth and first order refer to an expansion in the scattering potential. All the spin
  figures are normalized to the average value per conduction electron,
  $S=\frac{1}{2}$. The domain wall length is $\lambda=10\lambda_F$
  with (a) $\frac{J}{2}=0.112$~eV and $T=25$K, and (b)
  $\frac{J}{2}=2.23$~eV and $T=50$K.}\label{spinzc}
\end{figure}
\begin{figure}
  \includegraphics*[width=0.47\textwidth]{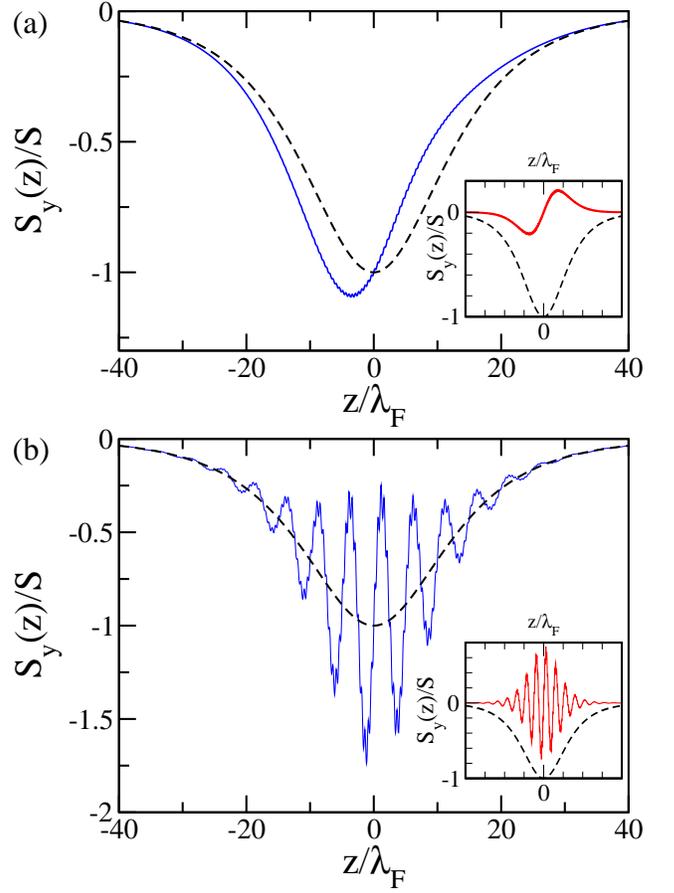}
  \caption[Spin Density]{(Color online) Same as Fig.~\ref{spinzc} but
    for the spin density $S_{y}(z)/S$.}\label{spinyc}
\end{figure}
In the following, we want to consider two examples. Example (a)
corresponds to a case where $\lambda_+ \ll \lambda < \lambda_-$ so
that only forward scattering contributes. In example (b), on the other
hand, we consider the case $\lambda_+\ll \lambda_- \lesssim \lambda$
so that forward scattering is again dominant but is oscillating over
the length of the domain wall. More specifically, we consider values
which are appropriate for Fe, $\lambda_F=0.367$ nm, and a lattice
cutoff $\alpha=\lambda_F$.  We plot results for two sets of
parameters: (a) $\frac{J}{2}=0.112$~eV, $\lambda=10\lambda_F$, and
$T=25$K; and (b) $\frac{J}{2}=2.23$~eV, $\lambda=10\lambda_F$, and
$T=50$K. These parameters give us the following scaling terms: (a)
$\gamma_f=1.12$, and $\gamma_b=0.923$; and (b)
$\gamma_f=1.09$, and $\gamma_b=0.937$. In both cases
the conditions \eqref{famp} and \eqref{bamp} are fulfilled with the
ratio of r.h.s divided by l.h.s being of the order $10^{-3}$ for the
condition on backward scattering, Eq.~\eqref{bamp}, and of the order
$10^{-1}$ for forward scattering, Eq.~\eqref{famp}. 

Figs.~\ref{spinzc}, and \ref{spinyc} show the spin density around the
domain wall, normalized to the average value per conduction electron,
$S=\frac{1}{2}$. For the situation where the length-scale of the
forward scattering oscillations is larger than the domain wall length,
case (a), the spin density profile is significantly altered, see
Figs.~\ref{spinzc}(a), \ref{spinyc}(a). Such a shift will affect how
the domain wall itself behaves in the effective magnetic field applied
by the conduction electrons and therefore will strongly affect the
domain wall dynamics.  As expected, the backward scattering plays no
role in the considered temperature range.

The asymmetric distortion of the spin density clearly visible in
Figs.~\ref{spinzc}(a) and \ref{spinyc}(a) is due to the addition of
antisymmetric and symmetric combinations of the spin densities.  In
contrast to Figs.~\ref{spinzc}(b) and \ref{spinyc}(b), where the
Friedel oscillations are rapid when compared to the domain wall
length, here the Friedel oscillations from forward scattering are on a
longer length-scale than $\lambda$. Hence the changes in the spin
density they cause can be seen as an overall distortion in its
profile.

When both $\lambda_+$ and $\lambda_-$ are smaller than the domain wall
length, see Figs.~\ref{spinzc}(b), \ref{spinyc}(b), oscillations
within the overall domain wall profile are clearly visible. Here the
long wavelength oscillations are caused by forward scattering while
the much faster oscillating backward scattering term causes the small
``wiggles'' on top of the oscillations. In this case the overall spin
density profile is not shifted.

\begin{figure}
\includegraphics*[width=0.45\textwidth]{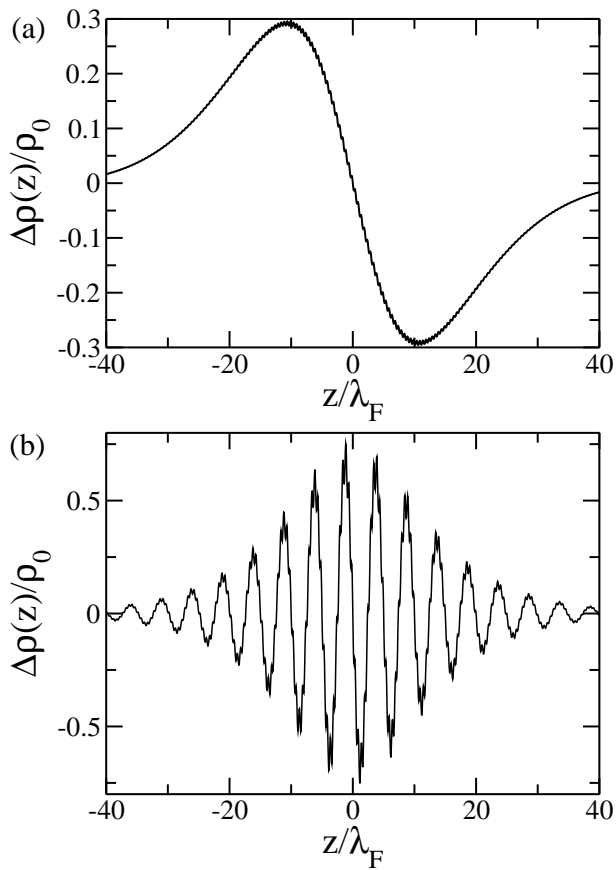}
\caption[Charge Density]{The charge density correction,
  $\Delta\rho(z)$, per electron per unit cell:
  $\rho_0=2e/\alpha$Cm${}^{-1}$. The domain wall length is
  $\lambda=10\lambda_F$ with parameters for panels (a) and (b) as
  given in Fig.~\ref{spinzc}.}\label{chargec}
\end{figure}

\begin{figure}
\includegraphics*[width=0.45\textwidth]{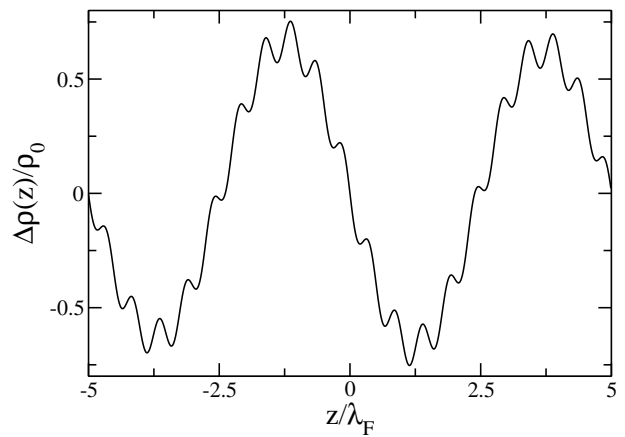}
\caption[Charge Density]{
  Same as Fig.~\ref{chargec}(b) where we have zoomed in on a shorter
  length making the oscillations due to back-scattering clearly
  visible.}\label{charges}
\end{figure}

The first order charge density correction, derived similarly, is given
by
\begin{eqnarray}
\langle \Delta \rho(z)\rangle&=&e\bigg\langle-\sqrt{\frac{2}{\pi}}\partial_z\phi_c(z)\\&&\nonumber+\frac{1}{\alpha\pi}\sum_{\sigma}\cos[2zk_{F\sigma}-2\sqrt{\pi}\phi_\sigma(z)] \bigg\rangle.
\end{eqnarray}
The main contribution to the charge density can be calculated
analytically and can be found in the appendix, Eq.~(\ref{rho}).
Results for the cases (a) and (b) are shown in Fig.~\ref{chargec}.
In case (a) we do see a charge build up, and respectively depletion,
antisymmetric with respect to the center of the domain wall. The small
``wiggles'' on top of the overall charge rearrangement are caused by
backward scattering. In case (b) we see strong oscillations of the
charge density caused by forward scattering which are largest at the
center of the domain wall. The very fast oscillations, which can be
seen in more detail in Fig.~\ref{charges}, originate from back
scattering.
Finally, we note that the overall charge in the model is conserved,
i.e., $\int \ud z\Delta\rho(z)=0$. The total spin $\langle S^2\rangle$
in the system is also conserved and is related to the charge density:
$\langle S^2(z)\rangle=\frac{3}{4e}\langle\rho(z)\rangle$. Therefore
the total spin is redistributed throughout the wire in precisely the same way as
the charge. However none of $S^x$, $S^y$, or $S^z$ are themselves
conserved.

If we compare Fig.~\ref{chargec} to the results found by Dugaev et
al., Ref.~\onlinecite{PhysRevB.65.224419}, the most striking
difference is the presence of strong Friedel oscillations. Such
oscillations are well known and are already present in the
non-interacting case. However, we also see that depending on
parameters the amplitude of the Friedel oscillations can be quite
small (see e.g.~Fig.~\ref{chargec}(a)) and might be easily overlooked
if one mainly concentrates on the overall shape. 
Compared to the non-interacting case\cite{PhysRevB.79.212410} the spin
density corrections have a different profile around the domain wall in
the transverse direction.
We want to stress that our model offers a method of calculating {\it
  non-perturbatively} the full effect of short range electron-electron
interactions on any physical property we are interested in. Provided
of course that one is in the regime of sufficiently high temperatures
such that the perturbative analysis of the gauge potential remains
valid. This includes, in particular, the physically important regime
for quantum wires in ferromagnetic materials discussed above. In such
systems, however, the phases of the extended domain wall regime (see
table \ref{tab1}) should also be accessible.

We now want to briefly discuss consequences of the strong correlations
in the electronic systems for the dynamics of the domain wall. The
magnetization dynamics are described by the Landau-Lifschitz-Gilbert
equation, or some suitable generalization
thereof.\cite{LL9,Gilbert2004,PhysRevLett.93.127204} There are two
different aspects to this we must consider. One is the straightforward
point that the dynamics, \emph{over the length of the wall}, will be
affected by the different spin density of the Luttinger liquid
compared to the Fermi liquid or non-interacting case. The second, more
interesting point, is whether the derivation of the non-adiabatic
terms in the LLG equation are valid for a Luttinger liquid.

Following Zhang and Li\cite{PhysRevLett.93.127204} one can derive
contributions to the magnetization dynamics which allow for the fact
that the electrons do not instantaneously follow the magnetization
profile. One first writes a continuity equation for the spins,
assuming part of it to be always parallel to the bulk magnetization
and allowing a small deviation from this. In order to derive the
current dependent (so called $\beta$-) terms, those which drive the
domain wall along the wire, one assumes that
$j^{x,y,z}_s(z,t)=-\mu_BPj^{x,y,z}_e(z,t)n^{x,y,z}(z,t)/e$.
$\vec{j}_e(z,t)$ is the charge current and $P$ is the magnitude of the
polarization, whilst $\vec{j}_s(z,t)$ is the spin current. A quite
reasonable assumption in a Fermi liquid, this of course starts to look
more dubious in the case of a Luttinger liquid. In the standard
Luttinger liquid model spin and charge are of course uncorrelated and
possess different velocities. Thus this assumption would completely
fail. For us the situation is not so simple as we do not have
spin-charge separation, nonetheless what is obvious from our model is
that spin and charge are not fully correlated. One is forced to work
with the spin current and not the electric current and, as we have
already seen, the spin degrees of freedom can behave rather
differently for this model.

\section{Conclusion}
We have investigated a Luttinger liquid coupled to a non-collinear
ferromagnetic magnetization profile in the shape of a domain wall. The
domain wall acts as a spatially extended magnetic impurity for the
electrons and introduces both forward and backward scattering terms,
active over the length of the domain wall. In contrast to the well
studied case of point-like impurities in Luttinger liquids the finite
extent of the domain wall introduces a whole new layer of complexity
to the problem. In a renormalization group treatment of the scattering
terms, one has to distinguish between an extended regime and a sharp
regime at low temperatures where the domain wall effectively becomes a
delta function. In the extended regime the scattering terms are bulk
operators while they become boundary operators in the sharp regime. An
operator relevant in the extended regime can therefore become
irrelevant in the sharp regime.

Simplest to understand is the sharp regime. Here the spin-flip and
potential forward scattering terms are irrelevant or marginal,
respectively. The low-temperature fixed points are then determined by
the spin-flip and potential back scattering terms which both can be
either relevant or irrelevant. If both are irrelevant, then the fixed
point is the Luttinger liquid if potential back scattering is only
relevant for one spin channel with the other terms being irrelevant,
then the fixed point is an effective spinless Luttinger liquid. In all
other cases the fixed-point will be a spin and charge insulator.
Depending on the relative relevance of the backscattering operators,
the region of the insulator in the vicinity of the domain wall can
show a reduced, increased or unaffected polarization in comparison to
the bulk.

If we start with an extended domain wall, however, then the domain
wall length $\lambda$ will usually be large compared to the backward
scattering length $\lambda_+$, i.e., the backward scattering terms
will strongly oscillate over the length of the domain wall. This means
that the effective bare coupling---roughly proportional to the Fourier
mode of the domain wall potential commensurate with the backward
scattering oscillations---will be exponentially small
$\sim\exp(-\lambda/\lambda_+)$. So even if backward scattering is
relevant, temperatures $T/J<\exp(-\lambda/\lambda_+)$ are required in
order to make this scattering process important. The sharp regime is
therefore only physically relevant if we already start with a very
sharp domain wall (of the order of a few lattice sites) which could
possibly be realized by a nanoconstriction.

In the extended regime, the low temperature physics of the model is
not only determined by the relevance or irrelevance of the various
scattering terms but also by the hierarchy of the three different
length scales present in the problem. Analyzing the first order RG
equations for backward and forward scattering and taking the hierarchy
of the three length scales into account we could identify three
phases. If either, both the scattering terms are irrelevant, or they
are relevant but the associated length scales are much smaller than
the domain wall length $\lambda$, we find an adiabatic Luttinger
liquid. In this case the spins of the electrons follow the
magnetization profile and both charge and spin excitations are
gapless. If both scattering terms are relevant and their respective
length scales larger than $\lambda$ then the first order RG equations
suggest that the system will become a spin-charge insulator, i.e., the
domain wall will act as a perfectly reflecting barrier. This case
corresponds to the normal Kane-Fisher fixed point. 
Finally, there is the case of forward scattering being relevant and
the associated length scale being larger than $\lambda$, with backward
scattering being either irrelevant or having an associated length
scale which is smaller than $\lambda$. In this case the charge modes
are gapless and charge is allowed to pass through the domain wall
barrier. The spin modes, on the other hand, are locked to a specific
value over the length of the domain wall and the system becomes
insulating with respect to spin transport (C$1$S$0$ phase).
Physically, this means that electrons no longer undergo a change of
spin on passing through the domain wall. The latter phase has not been
discussed so far in this context and it is this phase which we believe
is most important in possible experimental realizations.

We also calculated the spin and charge densities around the domain
wall for physically reasonable parameters in a regime at high enough
temperatures so that even relevant scattering terms can be treated
perturbatively. Here one finds spin and charge distributions markedly
different to the previously reported mean field interaction case. Both
the overall profile, and the local distribution, of spin and charge
show different behaviour, including Friedel oscillations. A similar
result to ours for the lateral component of spin is found in the
non-interacting case,\cite{PhysRevB.79.212410} though the transverse
components look qualitatively different. As an outlook, we believe
that it will be interesting to study how the dynamics of the domain
wall is changed in this temperature range where correlation effects
dramatically alter the spin and charge densities compared to the
non-interacting case but where we are still far above the phase
transition temperatures to the low-temperature phases discussed above.

\section*{Acknowledgments}
The authors wish to thank both J.~Berakdar and R.G.~Pereira for useful
and stimulating discussions. This work was supported by the DFG via
the SFB/Transregio 49 and the MAINZ (MATCOR) graduate school of
excellence.

\appendix*

\section{Details of the Rotations}

The spin and charge Luttinger parameters are, to lowest orders:
\begin{eqnarray}
  v_c&=&\frac{v_{F\uparrow}+v_{F\downarrow}}{2}+\frac{g_{4\parallel\uparrow}+g_{4\parallel\downarrow}+2g_{4\perp}}{4\pi},\nonumber\\
  v_s&=&\frac{v_{F\uparrow}+v_{F\downarrow}}{2}+\frac{g_{4\parallel\uparrow}+g_{4\parallel\downarrow}-2g_{4\perp}}{4\pi},\nonumber\\
  K_c&=&1-\frac{1}{2\pi}\frac{g_{2\parallel\uparrow}+g_{2\parallel\downarrow}+2g_{2\perp}}{v_{F\uparrow}+v_{F\downarrow}},\nonumber\\
  K_s&=&1-\frac{1}{2\pi}\frac{g_{2\parallel\uparrow}+g_{2\parallel\downarrow}-2g_{2\perp}}{v_{F\uparrow}+v_{F\downarrow}},\nonumber\\
  v_a&=&\frac{v_{F\uparrow}-v_{F\downarrow}}{2}-\frac{g_{2\parallel\uparrow}-g_{2\parallel\downarrow}}{4\pi}+\frac{g_{4\parallel\uparrow}-g_{4\parallel\downarrow}}{4\pi}\textrm{, and}\nonumber\\
  v_b&=&\frac{v_{F\uparrow}-v_{F\downarrow}}{2}+\frac{g_{2\parallel\uparrow}-g_{2\parallel\downarrow}}{4\pi}+\frac{g_{4\parallel\uparrow}-g_{4\parallel\downarrow}}{4\pi}.\label{ks}
\end{eqnarray}
In order to enforce the condition $K_s=1$ at the $SU(2)$ symmetric point ($J=0$) we set
\begin{eqnarray}
  K_s&=&1-\frac{1}{2\pi}\frac{g_{2\parallel\uparrow}+g_{2\parallel\downarrow}}{v_{F\uparrow}+v_{F\downarrow}},
\end{eqnarray}
canceling the $g_{2\perp}$ term by hand. It is clear that this $J$
independent term must cancel when higher order corrections are
included. What we do not know in this low-order analysis is how the
$J$ dependence of the Luttinger parameters will be modified by higher
order terms.

The two rotations we use on the bosonic fields are
\begin{eqnarray}\label{u}
  \begin{pmatrix}\phi_\uparrow \\ \phi_\downarrow\\ \theta_\uparrow\\ \theta_\downarrow\end{pmatrix}=\underbrace{\frac{1}{\sqrt{2}}\begin{pmatrix}
-1 & 0  & -1 & 0 \\
0 & -1 & 0 & -1 \\
1&0 & -1 & 0 \\
0&1  & 0 & -1
\end{pmatrix}}_{=\mathcal{U}^{-1}}\begin{pmatrix}\phi_{\uparrow+} \\ \phi_{\downarrow+}\\ \phi_{\uparrow-}\\ \phi_{\downarrow-}\end{pmatrix}
\end{eqnarray}
and
\begin{eqnarray}\label{ut}
\begin{pmatrix}\phi_s \\ \phi_c\\ \theta_s\\ \theta_c\end{pmatrix}=\underbrace{\frac{1}{\sqrt{2}}\begin{pmatrix}
1 & -1  & 0 & 0 \\
1 & 1 & 0 & 0 \\
0&0 & 1 & -1 \\
0&0  & 1 & 1
\end{pmatrix}}_{=\tilde{\mathcal{U}}^{-1}}\begin{pmatrix}\phi_\uparrow \\ \phi_\downarrow\\ \theta_\uparrow\\ \theta_\downarrow\end{pmatrix}.
\end{eqnarray}

To diagonalize the matrix $\tilde M$, Eq.~\eqref{M_SpinCharge}, we
perform two steps. The first is a rotation to make the Hamiltonian
diagonal, characterized by the $R_{1,2}^{\theta,\phi}$ terms. The
second is a rescaling to leave us with two distinct eigenvalues rather
than four, this introduces the $\Gamma$'s. Together this gives us
\begin{eqnarray}
\begin{pmatrix}\theta_1(z)\\ \theta_2(z)\end{pmatrix}&=&\nonumber
\begin{pmatrix} R^{\theta}_{1}\sqrt{\Gamma_1^\theta} & R^{\theta}_{2}\sqrt{\Gamma_1^\theta} \\ -R^{\theta}_{2}\sqrt{\Gamma_2^\theta} & R^{\theta}_{1}\sqrt{\Gamma_2^\theta} \end{pmatrix}
\begin{pmatrix}\theta_c(z)\\ \theta_s(z)\end{pmatrix}\textrm{ and}\\\begin{pmatrix}\phi_1(z)\\ \phi_2(z)\end{pmatrix}&=&
\begin{pmatrix} R^{\phi}_{1}\sqrt{\Gamma_1^\phi} & R^{\phi}_{2}\sqrt{\Gamma_1^\phi} \\ -R^{\phi}_{2}\sqrt{\Gamma_2^\phi} & R^{\phi}_{1}\sqrt{\Gamma_2^\phi} \end{pmatrix}
\begin{pmatrix}\phi_c(z)\\ \phi_s(z)\end{pmatrix}.\label{rotate0}
\end{eqnarray}
Here the rotation components are
\begin{eqnarray}\label{rotate1}
2(R^{\theta}_{1,2})^2&=&1\pm\bigg[1+\frac{v_a^2}{(\frac{v_cK_c}{2}-\frac{v_sK_s}{2})^2}\bigg]^{-\frac{1}{2}}\textrm{, and}\nonumber\\
2(R^{\phi}_{1,2})^2&=&1\pm\bigg[1+\frac{v_b^2}{(\frac{v_c}{2K_c}-\frac{v_s}{2K_s})^2}\bigg]^{-\frac{1}{2}}.
\end{eqnarray}
In order for the rotated Hamiltonian to have positive eigenvalues the
condition $v_cv_s/v_b^2>K_cK_s>v_a^2/v_cv_s$ must also be satisfied.
This condition seems to be always fulfilled, at least if one uses the
integrable Hubbard model as the underlying microscopic lattice
model.\cite{PhysRevB.47.6273}

The rescaling of the fields requires
\begin{eqnarray}
\label{A6}
(\Gamma_{1,2}^\phi)^2&=&(\Gamma_{1,2}^\theta)^{-2}\textrm{, and}\\
(\Gamma_{1,2}^\phi)^2&=&\frac{\frac{v_c}{2K_c}+\frac{v_s}{2K_s}\pm\sqrt{v_b^2+ (\frac{v_c}{2K_c}-\frac{v_s}{2K_s})^2}}{\frac{v_cK_c}{2}+\frac{v_sK_s}{2}\pm\sqrt{v_a^2+(\frac{v_cK_c}{2}-\frac{v_sK_s}{2})^2}}.\nonumber
\end{eqnarray}
Note that as the rotation is a unitary transformation we have the
condition $[R^{\theta,\phi}_1]^2+[R^{\theta,\phi}_2]^2=1$. For our
convenience we finally define
\begin{eqnarray}
\label{A7}
T^{\theta,\phi}_{1}&=&R^{\theta,\phi}_{2}/\sqrt{\Gamma_{1}^{\theta,\phi}},\\
T^{\theta,\phi}_{2}&=&R^{\theta,\phi}_{1}/\sqrt{\Gamma_{2}^{\theta,\phi}},\nonumber\\
\tilde{T}^{\theta,\phi}_{1}&=&R^{\theta,\phi}_{1}/\sqrt{\Gamma_{1}^{\theta,\phi}}\textrm{, and}\nonumber\\ \nonumber
\tilde{T}^{\theta,\phi}_{2}&=&- R^{\theta,\phi}_{2}/\sqrt{\Gamma_{2}^{\theta,\phi}}.
\end{eqnarray}
Therefore
the inverse of Eq.~\eqref{rotate0} is
\begin{eqnarray}
\begin{pmatrix}\phi_c(z)\\ \phi_s(z)\end{pmatrix}&=&
\begin{pmatrix} \tilde{T}^{\phi}_{1} & \tilde{T}^{\phi}_{2} \\ T^{\phi}_{1} & T^{\phi}_{2} \end{pmatrix}
\begin{pmatrix}\phi_1(z)\\ \phi_2(z)\end{pmatrix}\nonumber
\textrm{ and}\\
\begin{pmatrix}\theta_c(z)\\ \theta_s(z)\end{pmatrix}&=&
\begin{pmatrix} \tilde{T}^{\theta}_{1} & \tilde{T}^{\theta}_{2} \\ T^{\theta}_{1} & T^{\theta}_{2} \end{pmatrix}
\begin{pmatrix}\theta_1(z)\\ \theta_2(z)\end{pmatrix}.\label{rotate2}
\end{eqnarray}
Finally, the values of the new eigenvalues are
\begin{eqnarray}
u_{1,2}^2=\bigg(\frac{v_c}{2K_c}+\frac{v_s}{2K_s}\pm\sqrt{v_b^2+\big(\frac{v_c}{2K_c}-\frac{v_s}{2K_s}\big)^2}\bigg)\qquad\\ \times\bigg(\frac{v_cK_c}{2}+\frac{v_sK_s}{2}\pm\sqrt{v_a^2+\big(\frac{v_cK_c}{2}-\frac{v_sK_s}{2}\big)^2}\bigg).\nonumber
\end{eqnarray}
As expected this reduces directly to the spin and charge excitation
velocities in the absence of spin asymmetry. In such a case the
Hamiltonian is already diagonal and the above rotation is no longer
necessary.

The analytical result for the charge density correction is
\begin{eqnarray}\label{rho}
\langle \Delta \rho(z)\rangle&=&\frac{e\sech[z/\lambda]}{2\pi mT\alpha\lambda}[T^\phi_1T^\theta_1+T^\phi_2T^\theta_2]\times
\\&&\bigg[g_f\sin[z/\lambda_-] \left(\frac{T}{T^*_1}\right)^{\alpha_1}\left(\frac{T}{T^*_2}\right)^{\alpha_2}\nonumber\\&&\nonumber
+g_b\sin[z/\lambda_+]\left(\frac{T}{T^*_1}\right)^{\beta_1}\left(\frac{T}{T^*_2}\right)^{\beta_2}\bigg].
\end{eqnarray}


\end{document}